\documentstyle[12pt]{article}

\newcommand{\be}{\begin{equation}}
\newcommand{\ee}{\end{equation}}
\newcommand{\bea}{\begin{eqnarray}}
\newcommand{\eea}{\end{eqnarray}}
\newcommand{\nn}{\nonumber \\}
\newcommand{\p}[1]{(\ref{#1})}

\newcommand\al{\alpha}
\newcommand\bt{\beta}
\newcommand\de{\delta}
\newcommand\ga{\gamma}
\newcommand\ep{\epsilon}
\newcommand\Tb{\overline{T}}
\begin{document}
\begin{flushright}
hep-th/9809190\\
September 1998
\end{flushright}
\centerline{\large\bf Partial breaking $N=4$ to $N=2$:
hypermultiplet}
\centerline{\large\bf as a Goldstone superfield}

\vskip0.6cm
\centerline{\bf S. Bellucci,}

\vskip.1cm
\centerline{\it INFN - Laboratori Nazionali di Frascati,}
\centerline{\it P.O. Box 13, I-00044 Frascati, Italy}

\vskip.3cm
\centerline{\bf E. Ivanov, S. Krivonos}

\vskip.1cm
\centerline{\it Bogoliubov Laboratory of Theoretical
Physics, JINR,}
\centerline{\it 141 980, Dubna, Moscow Region,
Russian Federation}

\vskip.3cm
\centerline{\it Talk given at the XI International
Conference}
\centerline{\it PROBLEMS OF QUANTUM FIELD THEORY}
\centerline{\it In memory of D.I. Blokhintsev}
\centerline{\it Dubna (Russia), July 13-17, 1998}

\vskip.2cm
\centerline{\it and 32nd International Symposium
Ahrenshoop}
\centerline{\it on the Theory of Elementary Particles}
\centerline{\it Buckow (Germany), September 1-5, 1998}
\vskip.5cm

\begin{abstract}{\small
\noindent We describe the partial breaking of $N=1\;D=10$
supersymmetry
down to $(1,0)\;d=6$ supersymmetry within the non-linear
realization
approach. The basic Goldstone superfield associated with
this breaking
is shown to be the $(1,0)\;d=6$ hypermultiplet superfield
$q^{ia}$ subjected to a non-linear generalization of the
standard
hypermultiplet superfield constraint.
The dynamical equations implied by this constraint
are identified as
the manifestly worldvolume supersymmetric
equations of the Type I
super $5$-brane in $D=10$. We give arguments in favour
of existence
of the appropriate brane extension of off-shell
hypermultiplet
action in harmonic superspace. Some related problems,
in particular,
the issue of utilizing other $(1,0)\; d=6$
supermultiplets as
Goldstone ones, are shortly discussed.}
\end{abstract}

\vskip.5cm
\noindent{\bf 1.Introduction.} Spontaneous breakdown of
any global symmetry is accompanied by appearance
of Goldstone fields.
They have
quantum numbers of generators of spontaneosly broken symmetries
and, as
the most chracteristic feature, are transformed inhomogeneously
under the action
of these generators. Namely, their transformations start
with a pure shift
by the appropriate group parameter. A nice geometric meaning
of Goldstone
fields is revealed within the non-linear realizations theory
\cite{1} -
\cite{3}: they can be identified with parameters of the coset
$G/H$ of the
spontaneously broken symmetry group $G$ over its unbroken symmetry
subgroup $H$,
the vacuum stability subgroup. When $G$ is realized on the coset
manifold $G/H$
by left shifts as its group of motions, the coset coordinates,
Goldstone fields,
are transformed non-linearly and inhomogeneously under
the $G/H$ part of
symmetries and undergo linear rotations under the action of the
subgroup $H$.
The theory of non-linear realizations give general recipes
how to construct
invariant actions of Goldstone fields and their couplings to all
other, "matter"
fields. A remarkable property of such actions is that any
$H$-invariant
action of matter fields can be made $G$-invariant by switching on
proper couplings to the Goldstone fields. Any model with
a linear realization
of spontaneously broken symmetry can be rewritten in terms of
the fields of
the corresponding non-linear realization by means of
appropriate equivalence
redefinitions of the original fields. The text-book example of
theories based on non-linear realizations is provided
by non-linear
sigma models of spontaneously broken internal symmetries.

In case of ordinary, bosonic symmetry groups (e.g, the internal
symmetry
groups) the Goldstone fields are obviously bosonic.
Spontaneously broken
{\it supersymmetry} (SUSY) necessarily requires
{\it Goldstone fermions} to be
present among
the parameters of the corresponding cosets \cite{vak}. One more novel
feature of the supersymmetry case
% of the latter case
is extending of the notion of the vacuum stability subalgebra:
besides the generators yielding homogeneous rotations (e.g. Lorentz
group) it includes also those generators from the coset which have as
the associated parameters the space-time coordinates (e.g, translation
generators), or the superspace Grassmann coordinates, if some
supersymmetries remain unbroken\footnote{It is a generic feature
of non-linear realizations
of space-time (super)symmetries, i.e. those including the space-time
group of motion (e.g,, Poincar\'e group) as a subgroup \cite{{2},{3}}.}.
The rest of
the coset parameters are treated as Goldstone fileds (superfields)
given on this
space-time (superspace). The corresponding generators are genuine
spontaneously broken symmetry generators.

The case when all supersymmetries are spontaneously broken
is referred to as the {\it total} spontaneous breaking of SUSY.
The corresponding coset manifold is parametrized by the space-time
coordinates and Goldstone fermion fields defined on this space-time.
E.g., in the case of totally spontaneously broken
$N=1\; d=4$ Poincar\'e
SUSY \cite{vak}, the coset parameters are the Minkowski space
coordinates $x^m$ and the Goldstinos $\psi^\alpha (x),
\bar\psi^{\dot\alpha}(x)$.

The case of spontaneous {\it partial} breaking of global
supersymmetry (PBGS) is tantamount
to the situation when a part of SUSY generators remains unbroken.
Then, in the coset approach, one is led to associate with
these generators
Grassmann coordinates extending the space-time to a superspace of the
unbroken SUSY and to treat the coset parameters associated with
the genuine spontaneously broken symmetry generators as superfields
on this superspace, Goldstone superfields. E.g., in a generic case
of partial breaking of $N=2 \; d=4$ SUSY down to $N=1\; d=4$ SUSY
the coset is parametrized by $N=1$ superspace coordinates
$\{x^m, \theta^\alpha,\bar\theta^{\dot\alpha} \} \equiv \{X^M\}$
and Goldstone fermionic $N=1$ superfields $\Psi^\alpha (X),
\Psi^{\dot\alpha}(X)$ \cite{{bw},{bg1}}. The Goldstone fermionic field
comes out as the first component of such a superfield.

The study of partial breaking of $N=2\; d=4$ SUSY in the coset space
approach in refs. \cite{bw} - \cite{bg3} revealed a few
peculiarities of such theories.
\begin{itemize}
\item The treatment of fermionic Goldstone superfields as the basic
unconstrained ones does not lead to a self-consistent theory:
the $N=1$ Goldstone multiplet includes ghost degrees
of freedom \cite{bw}. A way out
was proposed in \cite{bg1}:
it consists in considering central charge-extended $N=2$ SUSY with
putting
the central charge generators into the coset. Then the basic Goldstone
superfields prove to be chiral bosonic $N=1$ superfields associated
with the
central charge generators. The fermionic Goldstone superfields are
expressed as $N=1$ spinor derivatives of the basic ones by imposing
some covariant constraints on the relevant Cartan 1-forms (the so
called
inverse Higgs effect \cite{invh}). Actually, $N=1$ chirality of the
central charge Goldstone superfields is also one of the consequences
of the inverse Higgs constraints: these superfields are
originally introduced as general $N=1$ ones. For the
basic Goldstone superfields in \cite{bg1}, \cite{bg3} there was
obtained a
self-consistent $N=2$ invariant action with a non-linearly realized
second supersymmetry.

\item There exist several inequivalent $N=1$ Goldstone supermultiplets
related to the partial breaking $N=2 \rightarrow N=1$. The Goldstone
fermionic field can be embedded into different $N=1$ multiplets:
chiral \cite{bg1}, vector \cite{bg2} and tensor ones \cite{bg3}. These
versions correspond to different theories. Moreover,
it seems that they require to choose {\it different} central
(or semi-central) extensions of standard $N=2$ SUSY
as inputs for a non-linear realization.

\item The $N=1$ superfield Goldstone actions for all these versions
can be
treated as gauge-fixed forms of the world-volume superfield actions
of some
BPS superbranes, along the line of refs. \cite{{hp},{hlp}}
(see also, e.g.,
\cite{{agit},{iak}}). The $N=1$ chiral Goldstone superfield action
is recognized as that of the Type I super 3-brane in a flat $D=6$
background (the
action possesses the whole set
of $(1,0)\; D=6$ SUSY symmetries including the $D=6$ Lorentz symmetry;
all these symmetries except for the worldvolume $N=1\; d=4$ SUSY
are realized non-linearly, in a Goldstone fashion).
Even more interestingly, the $N=1$ vector
Goldstone multiplet action describes a super D3-brane and yields
the Born-Infeld action for the gauge vector field.

\item In accord with the general property of non-linear realizations
mentioned in the beginning, one can promote different $N=1$ matter
actions
to $N=2$ supersymmetric ones by coupling the former to Goldstone
superfields.
\end{itemize}

All the actions presented in \cite{{bg1},{bg2},{bg3}} are nonlinear,
"brane" generalizations of various familiar off-shell $N=1$ superfield
actions. On the other hand, there exists a good off-shell description
of theories with {\it linearly} realized $N=2\;  d=4$ SUSY, e.g.
in harmonic $N=2$ superspace \cite{gikos}. Then a natural question
arises: whether some of these theories can be promoted to those with
non-linearly realized higher SUSY, say $N=4$ SUSY, by constructing
the formalism of partial breaking of this higher SUSY down to
$N=2$ SUSY and identifying some of well-known $N=2$ superfields as
the Goldstone ones accompanying this breakdown \footnote{In a different, 
supergravity and string context with a {\it linear} realization of 
$N=4$ SUSY the partial breaking $N=4 \rightarrow N=2$ 
was discussed in \cite{kk}.}. Related questions are
as to what kind of superbranes could be associated with such theories, whether
a brane generalization of the harmonic analyticity underlying ordinary
$N=2$ theories exists, how many different Goldstone
$N=2$ multiplets are possible, etc.

In this talk we partly answer some of these questions and make some
proposals concerning other ones. We show that
%the existence of brane generalization of the $q$-hypermultiplet
the hypermultiplet superfield can be regarded as a Goldstone superfield.
It realizes the partial breaking of
$N=1\; D=10$ SUSY (amounting to properly central-charge extended
$N=4$ SUSY in $d=4$ or $(1,1)$ SUSY in $d=6$) down to $(1,0)\; d=6$ SUSY.
Using the coset space techniques, we present the explicit form of
non-linear
transformations of hidden symmetries, show that all the superfield coset
parameters are covariantly expressed through the hypermultiplet
Goldstone superfield and find a covariant nonlinear generalization of
the standard hypermultiplet constraint in ordinary $(1,0)\; d=6$
superspace
(or the central-charge extended $N=2\; d=4$ superspace)\cite{fs}.
We argue that the dynamical equation for the hypermultiplet Goldstone
superfield is a gauge-fixed form of the equations of motion of the
Type I super 5-brane in $D=10$ with manifest worldvolume $(1,0)\; d=6$
SUSY.
We also adduce arguments in favour of existence of the relevant
brane extension of harmonic analyticity and the harmonic superspace
off-shell hypermultiplet actions \cite{gikos}.

\vspace{0.4cm}
\noindent{\bf 2.$N=1, D=10$ Poincar\'e superalgebra in the
$d=6$ notation.}
Instead of dealing with $N=4$ and $N=2$ SUSY in $d=4$, we choose
as the starting point their higher-dimensional counterparts,
$N=1\; D=10$ and $(1,0)\; d=6$ Poincar\'e superalgebras.
Our basic reasoning is the desire to consider most symmetric
situation. The $d=4$ case can be then reproduced via dimensional
reduction. As we wish to construct a superfield description of partial
breaking of $N=1\; D=10$ SUSY down to $(1,0)\; d=6$ SUSY
\footnote{We could equally choose $(0,1)\; d=6$ SUSY subalgebra as
the unbroken one.}, it is natural to
write the full superalgebra in the $d=6$ notation. From the $d=6$
viewpoint
the $N=1\; D=10$ SUSY algebra is a sort of central-charge extended
$(1,1)$
Poincar\'e superalgebra. In the standard spinor notation (see, e.g.
\cite{hst} - \cite{sok6}) it is constituted by the following
set of generators
\be
N=1\; D=10\;\;\;\; SUSY \qquad \propto \quad
\left\{ Q^i_\alpha, P_{\alpha\beta}, S^{\beta a}, Z^{ia}
\right\}~, \label{setsus}
\ee
where
$$
\alpha, \beta = 1,...,4~, \quad i = 1,2~, \quad a = 1,2
$$
are, respectively, the $d=6$ spinor indices and the doublet indices
of two
commuting automorphism $SU(2)$ groups realized on the $Q$ and $S$
supertranslations generators. The basic anticommutation relations read
\be
\left\{ Q_{\al}^i,Q_{\bt}^j\right\}=\ep^{ij}P_{\al\bt}\; ,\quad
\left\{ Q_{\al}^i,S^{a\bt}\right\} = \de_{\al}^{\bt}Z^{ia}\; , \quad
\left\{ S^{i\al},S^{b\bt}\right\} = \ep^{ab}P^{\al\bt} \;. \label{susy}
\ee
The $d=6$ translation generator\footnote{We use the following
notation
$
A^{\al\bt}=\frac{1}{2}\ep^{\al\bt\ga\de}A_{\ga\de}~,\;\;
\ep_{\al\bt\ga\de}\ep^{\al\bt\ga\de}=24~,
$
$
V^{i}=\ep^{ij}V_j~, \quad \ep_{ik}\ep^{kj} = \delta^j_i~.
$
}
$P_{\alpha\beta} = -P_{\beta\alpha}
= {1\over 2}\epsilon_{\alpha\beta\rho\lambda}P^{\rho\lambda}$, together
with the "semi-central charge" generator $Z^{ia}$, form the $D=10$
translation generator.

To the generators \p{setsus} one should also add the generators of
the $D=10$ Lorentz
group $SO(1,9)$ which in the $d=6$ notation are naturally divided
into the following set
\be
SO(1,9) \qquad \propto \quad \left\{ M_{\al\bt\;\ga\de},\;\;
T^{ij}, \;\; \Tb{}^{ab}, \;\; K_{ia}^{\al\bt}\right\}~. \label{so19ge}
\ee
Here the generators $M$ and $T$ generate mutually commuting
$d=6$ Lorentz
group $SO(1,5)$ and the automorphism (or $R$-symmetry)
group $SO(4)\sim SU(2)\times SU(2)$, the generators
$K$ belong to the coset $SO(1,9)/SO(1,5)\times SO(4)$. The relevant
commutation relations are
\bea
&&\left[ M_{\al\bt\;\ga\de},M_{\al'\bt'\;\ga'\de'}\right]  =
 \ep_{\al\bt\;\al'\bt'}M_{\ga\de\;\ga'\de'}-
 \ep_{\ga\de\al'\bt'}M_{\al\bt\;\ga'\de'} - (\al'\bt' \leftrightarrow
\ga'\de') \nn
%-
%  \ep_{\al\bt\ga'\de'}M_{\ga\de\;\al'\bt'}+
%\ep_{\ga\de\ga'\de'}M_{\al\bt\;\al'\bt'} \; , \nn
&&\left[ M_{\al\bt\;\ga\de},K_{\al'\bt'}^{ia}\right]  =
   \ep_{\al\bt\al'\bt'}K_{\ga\de}^{ia}-
   \ep_{\ga\de\al'\bt'}K_{\al\bt}^{ia} \; , \nn
&&\left[ T^{ij},T^{kl}\right]  =  \ep^{ik}T^{jl}+\ep^{il}T^{jk}+
   \ep^{jk}T^{il}+\ep^{jl}T^{ik} \; , \nn
&&\left[ \Tb{}^{ab},\Tb{}^{cd}\right]  =  \ep^{ac}\Tb{}^{bd}+
  \ep^{ad}\Tb{}^{bc}+\ep^{bc}\Tb{}^{ad}+\ep^{bd}\Tb^{ac} \; , \nn
&&\left[ T^{ij},K_{\al\bt}^{ka} \right]  =
    \ep^{ik}K_{\al\bt}^{ja}+\ep^{jk}K_{\al\bt}^{ia} \; ,
\left[ \Tb{}^{ab},K_{\al\bt}^{ic} \right]  =
    \ep^{ac}K_{\al\bt}^{ib}+\ep^{bc}K_{\al\bt}^{ia} \; , \nn
&&\left[ K_{\al\bt}^{ia},K_{\ga\de}^{jb} \right]  =
  \ep^{ij}\ep^{ab}M_{\al\bt\;\ga\de}+\frac{1}{2}\ep_{\al\bt\ga\de}
\left(
   \ep^{ij}\Tb{}^{ab}+\ep^{ab}T^{ij}\right) \; , \label{so91} \\
&&\left[ M_{\al\bt\;\ga\de},P_{\al'\bt'}\right]  =
   \ep_{\al\bt\al'\bt'}P_{\ga\de}-
   \ep_{\ga\de\al'\bt'}P_{\al\bt} \; , \nn
&&\left[ M_{\al\bt\;\ga\de},Q_{\mu}^i\right]=\frac{1}{2}\left(
  \ep_{\al\bt\ga\mu}\de_{\de}^{\nu}+\ep_{\al\bt\mu\de}\de_{\ga}^{\nu}-
  \ep_{\mu\bt\ga\de}\de_{\al}^{\nu}- \ep_{\al\mu\ga\de}\de_{\bt}^{\nu}
   \right)Q_{\nu}^i \; , \nn
&&\left[ M_{\al\bt\;\ga\de},S^{a\mu}\right]=-\frac{1}{2}\left(
  \ep_{\al\bt\ga\nu}\de_{\de}^{\mu}+\ep_{\al\bt\nu\de}\de_{\ga}^{\mu}-
  \ep_{\nu\bt\ga\de}\de_{\al}^{\mu}- \ep_{\al\nu\ga\de}\de_{\bt}^{\mu}
   \right)S^{a\nu} \; , \nn
&&\left[ T^{ij},Q_{\al}^{k} \right]  =
    \ep^{ik}Q_{\al}^{j}+\ep^{jk}Q_{\al}^{i} \; , \quad
\left[ \Tb{}^{ab},S^{c\al} \right]  =
    \ep^{ac}S^{b\al}+\ep^{bc}S^{a\al} \; , \nn
&&\left[ T^{ij},Z^{ka} \right]  =
    \ep^{ik}Z^{ja}+\ep^{jk}Z^{ia} \; , \quad
\left[ \Tb{}^{ab},Z^{ic} \right]  =
    \ep^{ac}Z^{ib}+\ep^{bc}Z^{ia} \; , \nn
&&\left[ P_{\al\bt},K_{\ga\de}^{ia}\right]=\ep_{\al\bt\ga\de}Z^{ia}\;,
       \quad
\left[ Q_{\al}^i, K_{\bt\ga}^{ja}\right]=\ep^{ij}\ep_{\al\bt\ga\de}
      S^{a\de}\; \nn
&&\left[ S^{a\al},K_{\bt\ga}^{ib}\right] = \ep^{ab}\left(
    \de_{\bt}^{\al}Q_{\ga}^i-\de_{\ga}^{\al}Q_{\bt}^i\right) \; ,
\quad
\left[ Z^{ia},K_{\al\bt}^{jb}\right] = \ep^{ij}\ep^{ab}P_{\al\bt}
     \; . \label{cos}
\eea

\vspace{0.4cm}
\noindent{\bf 3. Coset space and transformations.} We are going to
construct a non-linear realization of $N=1\; D=10$ SUSY (together with
the $D=10$ Lorentz group), such that $(1,0)\; d=6$ SUSY remains
unbroken.
So, following the generic coset approach prescriptions, we are led
to choose the vacuum stability subgroup to be
\be
H \quad \propto \quad \left\{Q^i_\alpha, P_{\alpha\beta}, T^{(ij)},
T^{(ab)},
M_{\al\bt\;\ga\de}
\right\}~. \label{Hsub}
\ee
We included into $H$ the maximal subgroup of $SO(1,9)$ with respect to
which the algebra of $Q^i_\alpha, P_{\alpha\beta}$ is closed, namely
$\tilde{H}= SO(4)\times SO(1,5) \propto \left\{M_{\al\bt\;\ga\de},
T^{(ij)}, T^{(ab)}\right\}$. This subgroup will produce purely
homogeneous rotations of all
involved objects and it is the genuine linear subgroup from
the standpoint
of the coset manifolds approach. Then we put the generators $Q^i_\alpha,
P_{\alpha\beta}$ into the coset and associate with them as the coset
parameters the coordinates of $(1,0)\; d=6$ superspace
\be
Q^i_\alpha \Rightarrow \theta^\alpha_i~, \quad P_{\alpha\beta}
\Rightarrow x^{\alpha\beta}~. \label{SS}
\ee
The remaining coset generators, $S^{\alpha a}, Z^{ia},
K^{ia}_{\alpha\beta}$, correspond to genuine spontaneously
broken symmetries
and the corresponding coset parameters are Goldstone superfields
on the $(1,0)$ $d=6$ superspace $\{x^{\alpha\beta},
\theta^\alpha_i \}$
\be
S^{\alpha a} \Rightarrow \Psi_{\alpha a}(x, \theta)~, \quad
Z^{ia} \Rightarrow q_{ia}(x,\theta)~, \quad K^{ia}_{\alpha\beta}
\Rightarrow \Lambda^{\alpha\beta}_{ia}(x,\theta)~. \label{Golddef}
\ee

As the next step, one should choose the appropriate parametrization of
the element $g$ of the coset space $G/\tilde{H}$ where $G$ is the full
supergroup of $N=1\; D=10$ SUSY, including the $D=10$ Lorentz group.
We use the exponential parametrization
\be
g=e^{x^{\al\bt}P_{\al\bt}}e^{\theta^{\al}_iQ_{\al}^i}e^{q_{ia}Z^{ia}}
  e^{\Psi_{a\al}S^{a\al}}e^{\Lambda_{ia}^{\al\bt}K_{\al\bt}^{ia}} \;.
\label{g}
\ee

Acting on \p{g} from the left by different elements of $G$ with
constant parameters, one can determine the transformation properties
of the coset coordinates and superfields.

Unbroken supersymmetry $(g_0=\mbox{exp }(a^{\al\bt}P_{\al\bt}+
  \eta_i^{\al}Q_{\al}^i ))$:
\be\label{susy1}
\de x^{\al\bt}=a^{\al\bt}+\frac{1}{4}\left(
\eta^{i\al}\theta_i^{\bt} - \eta^{i\bt}\theta_i^{\al} \right),\quad
\de \theta_{i}^{\al}=\eta_i^{\al}\; .
\ee

Broken supersymmetry $(g_0=\mbox{exp }(\eta_{a\al}S^{a\al}))$:
\be\label{susy2}
\de x^{\al\bt}=\frac{1}{4}\ep^{\al\bt\ga\de}\eta_{\ga}^a\Psi_{a\de},
\quad
\de q_{ia}=-\eta_{a\al}\theta_i^{\al},\quad
\de\Psi_{a\al}=\eta_{a\al}\; .
\ee

Broken $Z$-translations $(g_0 = \mbox{exp}(c_{ia}Z^{ia}))$:
\be \label{Ztr}
\delta q^{ia} = c^{ia}
\ee

Broken $K$ transformations $(g_0=\mbox{exp }(r_{ia}^{\al\bt}
K_{\al\bt}^{ia}))$:
\bea
&&\de x^{\al\bt}=-r^{ia\al\bt}q_{ia}-r_i^{a\ga\bt}\theta^{i\al}
  \Psi_{a\ga}+\frac{1}{2}\ep^{\al\bt\mu\nu}r_{\mu\ga}^{ia}
  \Psi_{a\nu} \; , \nn
&&\de \theta_i^{\al} = 2r_i^{b\al\bt}\Psi_{b\bt} \; , \nn
&&\de q_{ia} = -2r_{ia}^{\al\bt}x_{\al\bt}+r^j_{a\al\bt}\theta_i^{\al}
   \theta_j^{\bt}-r_i^{b\al\bt}\Psi_{b\al}\Psi_{a\bt} \; , \nn
&& \de \Psi_{a\al}=2r^i_{a\al\bt}\theta^{\bt}_i,\quad
   \de \Lambda_{ia}^{\al\bt}=r_{ia}^{\al\bt}+\ldots \quad .
   \label{ktr}
\eea

As was already mentioned, the subgroup
$\tilde{H} = SO(1,5)\times SO(4)$
is realized as rotations of the $SO(1,5)$ spinor and $SU(2)$
doublet indices.

We see that the $N=1\; D=10$ supergroup as a whole admits
a realization on
the coordinates of $(1,0)\; d=6$ superspace and Goldstone
superfields
"living" on this superspace. It is easy to check that the closure of
the above infinitesimal transformations is just the $N=1\; D=10$
superalgebra presented in the previous Section.

\vspace{0.4cm}
\noindent{\bf 4. Cartan forms.} Next and important step of the coset
approach is the construction of the Cartan 1-forms
which are used to define covariants of given non-linear realization.
They are defined by the generic formula
\be
g^{-1}d g = \Omega_Q + \Omega_P + \Omega_Z + \Omega_S + \Omega_K +
\Omega_{\tilde H}~,
\ee
with
\bea
\Omega_Z & \equiv & \Omega_Z^{ia}\;Z_{ia} = \left[ \left( \mbox{ch }
\sqrt{\varphi}\right)^{ia}_{jb}
   d{\hat q}^{jb}+\left( \frac{\mbox{sh }\sqrt{\varphi}}{\sqrt{\varphi}}
\right)^{ia}_{jb}2\Lambda^{jb\mu\nu}d{\hat x}_{\mu\nu} \right]Z_{ia} \nn
\Omega_P & \equiv & \Omega_P^{\alpha\beta}\;P_{\alpha\beta} =
\left[ \left( \mbox{ch } \sqrt{\phi}
      \right)_{\mu\nu}^{\al\bt}
   d{\hat x}^{\mu\nu}+\left( \frac{\mbox{sh }\sqrt{\phi}}{\sqrt{\phi}}
\right)_{\mu\nu}^{\al\bt}\Lambda_{ia}^{\mu\nu}d{\hat q}^{ia} \right]
  P_{\al\bt} \nn
\Omega_Q & \equiv & -\Omega_Q^{i\alpha}\; Q_{i\alpha} =
\left[ -\left( \mbox{ch } \sqrt{v} \right)^{i\al}_{j\bt}
   d{\theta}^{j\bt}-\left( \frac{\mbox{sh }\sqrt{\varphi}}
{\sqrt{\varphi}}
   \right)^{i\al}_{j\ga}2\Lambda^{ja\bt\ga} d \Psi_{a\bt} \right]
Q_{i\al}
   \nn
\Omega_S & \equiv & \Omega_{S a\beta}\;S^{a\beta} = \left[
\left( \mbox{ch }
\sqrt{\omega}
   \right)_{a\bt}^{b\ga}
   d{\Psi}_{b\ga}-\left( \frac{\mbox{sh }\sqrt{\omega}}
{\sqrt{\omega}}
\right)_{a\bt}^{b\ga}2\Lambda_{ib\al\ga}d{\theta}^{i\al}
\right]S^{a\bt}~.
\label{cforms}
\eea
Here
\bea
d{\hat x}^{\al\bt}& = & dx^{\al\bt}-\frac{1}{4}\theta^{i\al}
d\theta_i^{\bt}
  +\frac{1}{4}\theta^{i\bt}d\theta_i^{\al}-
  \frac{1}{4}\epsilon^{\al\bt\mu\nu}\Psi^a_{\mu}d\Psi_{a\nu} \nn
d{\hat q}_{ia}&=& dq_{ia}+\Psi_{a\al}d\theta_i^{\al}~,
\label{hat} \\
\varphi^{ia}_{jb} & \equiv & 2 \Lambda^{ia\mu\nu}
\Lambda_{jb\mu\nu} \; , \;
\phi_{\al\bt}^{\mu\nu} \equiv 2 \Lambda_{ia\al\bt}\Lambda^{ia\mu\nu}
\;,\nn
v^{i\al}_{j\bt} & \equiv & -4 \Lambda^{ib\ga\al}\Lambda_{jb\bt\ga}
\; , \;
\omega_{a\al}^{b\bt} \equiv -4 \Lambda_{ia\ga\al}\Lambda^{ib\bt\ga} \;.
\label{def2}
\eea
We do not give the explicit expressions for the coset Lorentz
form $\Omega_K =(d\Lambda^{\alpha\beta}_{ia} +...)K^{ia}_{\alpha\beta}$
and
the inhomogeneosly transforming form $\Omega_{\tilde H}$ on the
stability
subgroup as they are of no immediate relevance for our further
discussion.

\vspace{0.4cm}
\noindent{\bf 5. Inverse Higgs constraints and dynamical equation.}
By construction, the forms \p{cforms} are covariant under all
transformations of $G$ realized as left shifts of $g$. They merely
undergo some induced $\tilde H$ rotations in their spinor
and $SU(2)$ indices (these
rotations are field-dependent in the case of
the $G/\tilde{H}$-transformations).
This fact allows us to apply the inverse Higgs procedure to eliminate
all Goldstone superfields in favour of $q^{ia}(x,\theta)$.
Indeed, we observe that $\Psi_{\alpha a}$ and
$\Lambda^{\alpha\beta}_{kb} $ appear inside the form $\Omega_Z $
{\it linearly} as coefficients of the coordinate differentials
$d\theta^\beta_i$ and
$dx^{\alpha\beta}$, respectively. The Goldstone superfield
$\Lambda$
linearly
appears also in the form $\Omega_S$. Thus these superfields
are covariantly expressible in terms of $\theta $- and
$x$-derivatives of
$q^{ia}(x,\theta)$. The natural covariant constraint making this
job is as follows
\be
\Omega_Z = 0~. \label{basconstr}
\ee

It is easy to find that this constraint amounts to the following
set of equations
\bea
\partial_{\mu\nu}q_{ia}&=& \frac{1}{2}
 \left(\de_{\mu}^{\rho}\de_{\nu}^{\sigma}  -
  \de_{\mu}^{\sigma}\de_{\nu}^{\rho}
  -\frac{1}{2}
   \epsilon^{\rho\sigma\al\bt}\Psi_{\al}^b
\partial_{\mu\nu}\Psi_{b\bt}
    \right)\widetilde{\Lambda}_{ia\rho\sigma} \equiv
E^{\rho\sigma}_{\mu\nu}\;\widetilde{\Lambda}_{ia\rho\sigma} ,
\label{firste}
\\
{\cal D}_{\bt}^j q_{ia}-\de_i^j \Psi_{a\bt} & = &
 \frac{1}{4}\widetilde{\Lambda}_{ia\mu\nu}
   \epsilon^{\mu\nu\al\ga}\Psi_{\al}^b {\cal D}_{\bt}^j
\Psi_{b\ga} \; ,
\label{seconde}
\eea
where
\be
\widetilde{\Lambda}^{ia\mu\nu}\equiv
    -2\left( \frac{\mbox{th }\sqrt{\varphi}}{\sqrt{\varphi}}
\right)^{ia}_{jb} \Lambda^{jb\mu\nu}
\ee
and ${\cal D}_{\bt}^j$ is the ordinary flat $(1,0)\; d=6$
spinor derivative
\be
{\cal D}_{\bt}^j = {\partial \over \partial \theta^\bt_j} -{1\over 2}
\;\theta^{j\al}\partial_{\al\bt}~,\qquad \{{\cal D}^i_\al,
{\cal D}^k_\bt \} =
\epsilon^{ik}\partial_{\al\bt}~. \label{defD}
\ee
It is easy to directly check covariance of this system under, say,
the nonlinear supersymmetry transformations \p{susy2}.
Let us point out that there is actually no need to explicitly
check the covariance
as it directly stems from the manifest covariance of the constraint
\p{basconstr}.

Looking at the equations \p{firste}, \p{seconde} we observe that the
first equation and the trace part of the second one are indeed purely
algebraic nonlinear
relations allowing to trade $\widetilde{\Lambda} $ and $\Psi $ for
the $x$- and
$\theta$-derivatives of $q^{ia}$
\bea
&& \widetilde{\Lambda}_{ia\rho\sigma} =
(E^{-1})_{\rho\sigma}^{\mu\nu}
\;\partial_{\mu\nu}q_{ia}~, \label{lambdeq} \\
&& \Psi_{a\beta} = {1\over 2}\;\nabla_\beta^k \;q_{ka}~,
\label{psieq}
\eea
where
\be
\nabla^k_\beta \equiv {\cal D}^k_\bt -
{1\over 4}\;(E^{-1})^{\mu\nu}_{\rho\lambda}\;
\epsilon^{\rho\lambda\al\ga}(\Psi_{\al}^b {\cal D}_{\bt}^k \Psi_{b\ga})
\;\partial_{\mu\nu} = {\cal D}^k_\bt - {1\over 4}\;
\epsilon^{\mu\nu\al\ga}(\Psi_{\al}^b \nabla_{\bt}^k \Psi_{b\ga})
\;\partial_{\mu\nu}
\label{defnabl}
\ee
(for the time being, we are not aware of the full expression of
$\Psi_{\al}^b$
through $q^{ia}$, only a few first terms in the iteration solution of
\p{psieq} were found).

The remaining, isotriplet part of \p{seconde} yields
the following constraint on $q^{ia}$:
\be
\nabla^{(i}_\bt \;q^{k)a} = 0~. \label{basconstr2}
\ee
We recognize it as a nonlinear generalization of the well-known
hypermultiplet constraint
\cite{fs}
\be
{\cal D}^{(i}_\bt\; q^{k)a} = 0~. \label{constrfree}
\ee
It is known that it reduces the field content of
$q^{ia}(x, \theta)$ to
four bosonic and eight fermionic components
\be
q^{ia}(x, \theta) \;\;\Rightarrow \;\; \phi^{ia}(x) +
\theta^{\alpha i}
\psi_\alpha^a (x) + x\mbox{-derivatives}~, \label{reduct}
\ee
and simultaneously puts these fields on shell
\be
\Box \phi^{ia}(x) = 0~, \quad \partial^{\alpha\beta}\psi^a_\bt = 0
\quad
\left(\Box \equiv \partial^{\alpha\beta}\partial_{\alpha\beta}
= {1\over 2}
\;\epsilon^{\al\bt\mu\nu}\partial_{\alpha\beta}\partial_{\mu\nu}~,
\;\;\partial_{\alpha\beta}\partial^{\lambda\bt} =
{1\over 4}\;\delta^\lambda_\al\;\Box \right)~.
\ee
Eq. \p{basconstr2} is expected to yield a non-linear generalization of
the $d=6$ hypermultiplet irreducibility conditions and equations of
motion.

Inspecting how the spontaneously broken nonlinear (super)symmetries
\p{susy2} - \p{ktr} are realized on the components of $q^{ia}$
(at the linearized
level), we conclude that $\phi^{ia}(x)$ and $\psi_\alpha^a(x)$ are just
Goldstone fields associated with the broken $Z$-translations and $S$
-supertranslations, while the Goldstone fields accompanying
the spontaneous
breakdown of the $SO(1,9)/ SO(1,5)\times SO(4)$ transformations,
$\partial_{\alpha\beta}\phi^{ia}(x)$, are recognized as the
coefficients
of the second-order $\theta$ monomials in the $\theta$-expansion
of $q^{ia}(x,\theta)$.

The conclusion is that the only essential Goldstone superfield
supporting
the partial spontaneous breaking of $N=1\; D=10$ SUSY down to
$(1,0) \; d=6$
SUSY within the non-linear realization scheme is the hypermultiplet
superfield $q^{ia}(x, \theta)$. It is subjected to the nonlinear
dynamical constraint \p{basconstr2} and accomodates all the Goldstone
fields associated with the spontaneosly broken symmetry generators
including parameters of the $D=10$ Lorentz group
coset $SO(1,9)/ SO(1,5)\times SO(4)$.

Note that the kinematical eq. \p{psieq} and the dynamical eq.
\p{basconstr2} are separately covariant with respect to hidden
symmetries. This reflects the fact that they can be re-obtained from
equating to zero, respectively, the separately covariant
isosinglet and isotriplet
parts of the full covariant spinor derivative $\Delta^k_\nu q^{ia}$.
This derivative (together with the full covariant
vector derivative $\Delta_{\mu\nu}q^{ia}$)
is defined by the standard relation
\be
\Omega_Z^{ia} \equiv \Omega_P^{{}\mu\nu}\;(\Delta_{\mu\nu}q^{ia}) +
\Omega_{Q\;k}^{\;\;\;\mu}\; (\Delta^k_\mu q^{ia})~.
\ee
Putting to zero the singlet part has the standard inverse Higgs
motivation as the condition of the covariant elimination of the
Goldstone spinor superfield $\Psi_{a\beta}$ in favour of $q^{ia}$.
However, the same requirement for the triplet part
(which produces the reduction of the field content of $q^{ia}$
and simultaneously yields the dynamics) has no such a clear
interpretation. It is not implied by the formalism of non-linear
realizations, and should be regarded as a kind of dynamical postulate.
In the superembedding approach to superbranes
(initiated in \cite{khark}, \cite{hs}) a similar postulate is known as
the "geometro-dynamical" principle
(see \cite{hs} and references therein)\footnote{It is worth noting
that in one of the first papers where
this postulate was introduced and exploited \cite{volzhel}
it was regarded as a sort of
inverse Higgs effect.}. An interplay between
the superembedding approach and the non-linear realizations
PBGS approach is discussed, e.g., in a recent preprint
\cite{goteb}. We will
return to this point in the concluding section.

For further discussion it will be convenient to project all
the involved
quantities on the $SU(2)$ harmonics $u^{\pm i}, u^{+i}u^-_i =1$
\cite{gikos}
\be
\theta^\al_i \;\Rightarrow \; \theta^{\pm \al} =
\theta^{\al i}u^\pm_i~, \;\;
{\cal D}^i_\al \;\Rightarrow \;{\cal D}^\pm_\al
= {\cal D}^i_\al u^\pm_i~,
\;\;\{D^+_\al, D^-_\beta \} = -\partial_{\al\beta}~, \;\;
q^{ia} \; \Rightarrow \; q^{\pm a} = q^{ia}u^\pm_i~.
\label{proj}
\ee
Then eqs. \p{basconstr2}, \p{psieq} can be written in the following
concise form
\bea
&& \nabla^+_\al q^{+a} = 0~,  \label{basconstr3} \\
&& \Psi^a_\beta =
- \nabla^-_\beta q^{+a} = \nabla^+_\beta q^{-a}~. \label{psieq3}
\eea

The covariant derivatives $\nabla^{\pm}$ satisfy the following
algebra
\bea
\{\nabla^+_\al, \nabla^-_\beta \} &\equiv &-\nabla_{\al\beta} =
-F^{\rho\lambda}_{\al\beta}\;\partial_{\rho\lambda}~, \label{+-} \\
\{\nabla^+_\al, \nabla^+_\beta \} &\equiv &-\nabla^{++}_{\al\beta} =
-F^{++\rho\lambda}_{\;\;\;\al\beta}\;\partial_{\rho\lambda}~,
\label{++} \\
F^{\rho\lambda}_{\al\beta} &=&
(E^{-1})^{\rho\lambda}_{\omega\sigma}\left[{1\over 2}\;(
\delta^\omega_\al \delta^\sigma_\beta -
\delta^\sigma_\al\delta^\omega_\beta)
+ \epsilon^{\omega\sigma\gamma\tau}(\nabla^+_\al\Psi^d_\gamma)\;
(\nabla^-_\beta\Psi_{d\;\tau}) \right]~,
\label{F0} \\
F^{++\;\rho\lambda}_{\;\;\;\al\beta} &=&
(E^{-1})^{\rho\lambda}_{\omega\sigma}
\epsilon^{\omega\sigma\gamma\tau}(\nabla^+_\al\Psi^d_\gamma)\;
(\nabla^+_\beta\Psi_{d\;\tau})~.
\label{F++}
\eea
The remaining anticommutator $\{\nabla^-_\al, \nabla^-_\beta \}$
follows from
\p{++}, \p{F++} via the replacement of indices $+ \rightarrow -$.

We observe that the derivative $\nabla_{\al\beta}$, alongside
with the $SU(2)$
singlet part, which starts with $\partial_{\al\beta}$
and is antisymmetric
in spinor indices, contains also a non-standard tripet part which
starts with a three-linear term and is {\it symmetric} in indices
$\al $, $\beta $. Just this second part appears in the r.h.s. of the
anticommutator \p{++}. This last property, at first sight, seems
to obscure the consistency of the dynamical constraint \p{basconstr3}
since it leads to the integrability condition
\be
\nabla^{++}_{\al\beta}q^+_a = 0~, \label{intcond}
\ee
which could be too strong. E.g., it could imply $q^{ia}$
to be constant.
However, we have checked that, up to seventh order in $q^{ia}$,
this condition is satisfied {\it identically} as a consequence
of the structure of $\nabla^{++}_{\al\beta}$. Though we are still
unable to prove this property in general, in what follows we take
for granted that \p{intcond} produces no new dynamical restrictions on
the superfield $q^{ia}$ (or $q^{+a}$).

Using the algebra \p{+-} - \p{F++} and eqs. \p{basconstr3}, \p{psieq3}
it is easy to find
\be
\nabla^+_\al \Psi_{a\beta} = -\nabla_{\al\beta}q^+_a~, \quad
\nabla^-_\al \Psi_{a\beta} = -\nabla_{\beta\al}q^-_a~. \label{nablpsi}
\ee
{}From the structure of covariant derivatives one immediately
concludes that
all superfields obtained by successive action of $\nabla^\pm_\al$
on $\Psi_{a\beta}$ are reduced to ordinary $x$-derivatives
of $q^{ia}$ and $\Psi_{a\beta}$, i.e. these two superfield projections
indeed exhaust the irreducible field content of $q^{ia}(x,\theta)$.

In the standard free hypermultiplet case an analog of the constraint
\p{basconstr3} reads (cf. \p{constrfree})
\be
{\cal D}^+_\beta q^{+a} = 0~. \label{constrfree1}
\ee
Hitting it by three appropriate ${\cal D}$'s and using their
anticommutator
algebra, one gets
\be
{\cal D}^+_\rho {\cal D}^-_\gamma {\cal D}^-_\nu {\cal D}^+_\beta q^+_a
= 0\;\;
\Leftrightarrow \;\;
\left(\partial_{\rho\gamma}\partial_{\nu\beta}
- \partial_{\gamma\beta}\partial_{\rho\nu} +
\partial_{\rho\beta}\partial_{\gamma\nu}\right)q^+_a = 0
\;\;\Leftrightarrow \;\; \Box \;q^+_a = 0~, \label{eqfree}
\ee
i.e. \p{constrfree1} puts $q^+_a$ on shell, in accord with
the said earlier. To find equations of motion in the full
nonlinear case, one can proceed in a similar way, replacing
${\cal D}^\pm $ by $\nabla^\pm $ in \p{eqfree}. Because of the
essential nonlinearity of the algebra \p{+-} - \p{F++}, the
analog of eq. \p{eqfree} looks rather complicated, but it
is simplified in the bosonic limit, when all fermionic components
are omitted
\be
\left(\nabla_{\beta\nu}\nabla_{\rho\gamma}
+ \nabla_{\nu\gamma}\nabla_{\rho\beta} -
\nabla_{\beta\gamma}\nabla_{\rho\nu}\right)q^+_a -
\{\nabla^+_\rho, [\nabla^-_\gamma, \nabla_{\beta\nu}]\}q^+_a
+ \{\nabla^+_\rho, [\nabla^+_\beta, \nabla_{\nu\gamma}]\}q^-_a = 0~.
\label{eqbos}
\ee
Note that the arrangement of indices in \p{eqbos} is important,
as the covariant vector derivatives do not commute with themselves and
with $\nabla^\pm_\al $. Besides, as we already mentioned,
they contain the parts both symmetric and antisymmetric in the
spinor indices. To see, what kind of dynamics
is hidden in \p{eqbos}, we considered it up to the first non-trivial
order in fields, the third order. Even in this lowest order the
calculations are rather tiresome though straightforward. We found
that it amounts to the following equation for $\phi^{ia}(x) \equiv
q^{ia}(x,\theta)|_{\theta = 0}$
\be
\Box \phi^{ia} +{1\over 2}\;\partial^{\rho\lambda}
\partial^{\mu\nu}\phi^{ia} \left(\partial_{\mu\nu}\phi \cdot
\partial_{\rho\lambda}
\phi\right) = 0~, \label{NGeq}
\ee
where we omitted three-linear terms containing $\Box $ as they
contribute to the next, 5th order, and used the notation
$A\cdot B \equiv A^{ia}B_{ia} $. All other terms which are present in
\p{eqbos} in this order and have more complicated $SU(2)$
representation content have been found either to
identically vanish or to contain terms $\sim \Box \phi^{kb}$.

Looking at \p{NGeq}, one observes that this equation just corresponds
to the "static gauge" form of the standard bosonic $5$-brane Nambu-Goto
action with the induced metric
\be
g_{\rho\lambda\;\; \mu\nu} =
{1\over 2}\;
\left( \epsilon_{\rho\lambda\mu\nu}
- \partial_{\rho\lambda}\phi \cdot\partial_{\mu\nu}\phi \right)
\equiv
{1\over 2}\;
\left( \epsilon_{\rho\lambda\mu\nu} - d_{\rho\lambda \;\;\mu\nu}
\right)~,
\label{metrind}
\ee
that is
\bea
S_{NG} &=& \mbox{const}\;\int d^6x \;\left(\sqrt{-\mbox{det}\;g}
-1\right) \nn
&\sim & \int d^6 x \left\{ \mbox{Tr}\;d -
{1\over 8}\left(\mbox{Tr}\;d\right)^2 + {1\over 4}\;\mbox{Tr}\;d^{\;4}
+ O(\phi^{\;6})\right\}. \label{NGact}
\eea

Though it still remains to prove that the higher-order
corrections are combined into this nice geometric form,
the above consideration suggests that this is very plausible.
Then eq. \p{basconstr3} should be interpreted as a
manifestly $(1,0)\; d=6$ world-volume superymmetric PBGS form of
the equations of the type I super 5-brane in $D=10$.
So the non-linear realization description of the partial breaking
of $N=1\; D=10$ SUSY down to $(1,0)\; d=6$ SUSY admits the
natural brane interpretation, much in line of the previous
studies \cite{bg1} - \cite{bg2}, \cite{goteb}.

Note that all the relations presented so far admit simple dimensional
reduction to the $d=5$ and further $d=4~, ..., 1$ worldvolumes by
neglecting dependence on the corresponding worldvolume coordinates.
Without entering into details, one gets in this way manifestly
worldvolume supersymmetric superfield equations describing super
4-brane
in $D=9$, super 3-brane in $D=8$ and so on, up to a superparticle in
$D=5$. In all these cases $8$ supersymmetries are realized linearly
in the relevant worldvolume superspaces, while the remaining $8$ are
realized nonlinearly.

\vspace{0.4cm}
\noindent{\bf 6. Brane extension of the off-shell $q^+$ action?}
In the case of ordinary hypermultiplet it is well-known
that just because the irreducibility constraint \p{constrfree}
implies equations of motion, no off-shell
superfield action exists for the hypermultiplet
in ordinary $(1,0)\; d=6$ ($N=2 \; d=4$) superspace. However,
the off-shell description becomes possible in the harmonic
superspace \cite{gikos}. There, eq. \p{constrfree} in the form
\p{constrfree1} is interpreted as
the analyticity condition implying that $q^+_a$ naturally "lives"
on some analytic subspace $\zeta^M = \{x^{\al\beta}_A,
\theta^{+\gamma},
u^\pm_i \}$ of the full harmonic $(1,0) \;d=6$ superspace, i.e.
$q^+_a \Rightarrow q^+_a (\zeta )$.
On the other hand, the homogeneity of $q^+$ in \p{proj} in harmonics
$u^+_i$ now follows from the equation of motion derived by the
analytic $q^+$ superfield action
\be
S_q = {1\over 2}\;\int d\zeta^{(-4)} q^{+a}{\cal D}^{++}q^+_a~.
\quad d\zeta^{(-4)}
\equiv d^6x_A d^4\theta^+[du]~. \label{qaction}
\ee
Here ${\cal D}^{++}$ is the analyticity-preserving harmonic
derivative in
the analytic basis of the harmonic $d=6$ superspace
\be
D^{++} = \partial^{++} -{1\over 2 }\;\theta^{+\al}\theta^{+\beta}
\partial_{\al\beta}~, \quad (\partial^{++} = u^{+i}\frac{\partial}
{\partial u^{-i}} )~.
\ee
The free $q^+$ action admits addition of self-interactions
which produce, in the generic case, a general hyper-K\"ahler
sigma model in the bosonic sector (with a four-dimensional
target in the case of one hypermultiplet).  Most characteristic
feature of the free off-shell action \p{qaction} and its
sigma model generalizations is the presence of
{\it infinite} sets of auxiliary fields.

One can wonder whether a brane generalization of eq.
\p{constrfree1}, i.e. eq. \p{basconstr3}, also admits an
interpretation as the analyticity condition and whether a brane
extension of the action \p{qaction} exists. By the brane
extension we understand the action such that the associated
equations of motion together with the analyticity conditions
amount to the basic dynamical constraint \p{basconstr3}.

The fact that the consistency condition \p{intcond} is satisfied
(at least up to seventh order, as we have checked) implies that
\p{basconstr3} indeed can be interpreted as a sort of Cauchy-Riemann
conditions defining a non-linear Grassmann harmonic analyticity
for $q^+_a$. Then, by analogy with
the standard hypertmultiplet case, this analyticity can be made
manifest by passing to a new basis in $(1,0)\; d=6$ harmonic
superspace where $\nabla^+_\al $ becomes "short" on $q^+_a$, i.e.
proportional to the partial derivative $\partial/ \partial
\theta^{-\al}$. Clearly, the relevant change of coordinates
should be highly nonlinear in $q^{ia}$ and its derivatives
(analogously to the relation between non-linear and manifest
$N=1$ chiralities in the case of the partial breaking
$N=2$ to $N=1$ \cite{bg1}). Unfortunately, for the time being
we do not know how to construct such nonlinear "bridges" within the
nonlinear realization formalism. One way is, of course, to
find them "by brute force", order by order in fields. But it seems
there exists another way around.

Namely, let us for a moment forget about eq. \p{basconstr3}
and deal with the manifestly analytic superfield $q^+_a (\zeta)$
having the free action \p{qaction}. In the bosonic sector, after
eliminating an infinite tower of auxiliary fields, it yields the
free action for the physical bosons $v^{ia}(x)$
($q^{+a} = v^{ia}(x)u^+_i \break +...$)
\be
S_q \quad \Rightarrow \quad {1\over 2}\;\int d^{\;6}x \;
\left(\partial v\cdot
\partial v \right)~.
\label{bosfree}
\ee
Assume that one succeeded in constructing a generalization
of \p{qaction},
such that in the bosonic sector it yields the $5$-brane Nambu-Goto
extension of \p{bosfree} in the form \p{NGact}. Then
this extension can naturally be expected to provide the analytic basis
description of the above "brane" hypermultiplet and to be the correct
Goldstone superfield action for the considered PBGS pattern.
All symmetries \p{susy1} - \p{ktr} found in the
central basis are expected to have their analytic basis counterparts
which play the decisive role in fixing the precise structure
of the "brane" $q^+$ action.

It is remarkable that the first simplest correction to \p{qaction}
which adds quadrilinear terms to the free bosonic action \p{bosfree}
arrange these terms just in the way required by the Nambu-Goto action!

This correction is almost uniquely fixed by the dimensionality
considerations
and the preservation of the harmonic $U(1)$ charge
\be
S_q \quad \Rightarrow \quad S_q + {\alpha \over 4}
\int d^6x_A d^8\theta [du]
\;(q^{+a}{\cal D}^{--}q^+_a)^2~.
\label{corr}
\ee
In the second term integration goes over
the {\it whole} $(1,0)\;d=6$ harmonic superspace, $\alpha $ is a
dimensionless parameter and
\be
{\cal D}^{--} = \partial^{--} -{1\over 2}\;\theta^{-\al}
\theta^{-\beta}
\partial_{\al\beta} +\theta^{-\al}{\partial\over
\partial \theta^{+\al}}~.
\ee
Passing to components and eliminating auxiliary fields (they
do not become propagating as one could anticipate) yield in the
fourth order in $v^{ia}(x)$ a few
terms of the fourth order in $x$ derivatives
\footnote{We normalize the Grassmann integrals over analytic
and full superspaces in the following way
$$\int d^4\theta^+ (\theta^+)^4 = \int d^4 \theta^+d^4\theta^-
(\theta^+)^4(\theta^-)^4 =1~, \qquad
(\theta^\pm)^4 \equiv {1\over 4!}\epsilon_{\al\beta\gamma\lambda}
\theta^{\pm\al}\theta^{\pm\beta}\theta^{\pm\gamma}
\theta^{\pm\lambda}~.
$$}. At first sight, these terms have nothing in common with
\p{NGact}: some of them involve a few derivatives on a single
$v^{ia}(x)$,
in such a way that they cannot be distributed as in \p{NGact}
by integrating by parts.

Surprisingly, these unwanted terms prove to be removable
by means of
appropriate non-linear redefinition of $v^{ia}(x)$:
they are cancelled by similar terms coming from the free
part of the action, i.e. \p{bosfree}. This change of variable
in the considered order in fields is
uniquely fixed by requiring such a cancellation:
\be
v^{ia} = \phi^{ia} +{\alpha\over 3}
\left\{\left[(\Box\phi\cdot \phi)
-(\partial\phi \cdot \partial\phi) \right]\phi^{ia}
-{1\over 4}\;(\phi)^2 \;\Box \phi^{ia} +{1\over 2}(\phi\cdot
\partial_{\mu\nu}\phi)\;\partial^{\mu\nu}\phi^{ia} \right\}
+ O(\phi^5)~.
\label{redef}
\ee
Finally, the bosonic part of the action \p{corr}, up to the fourth 
order in fields, acquires the following form
\be
S_b = {1\over 2}\int d^6x \left\{ (\partial\phi\cdot\partial \phi)
+{3\alpha \over 4}\left[
(\partial_{\mu\nu}\phi\cdot\partial_{\rho\lambda}\phi)
(\partial^{\rho\lambda}\phi\cdot\partial^{\mu\nu}\phi)
- {1\over 2}(\partial\phi\cdot \partial\phi)^2 + O(\phi^{\;6}) \right]
\right\}~.
\ee
It precisely coincides with \p{NGact} under the choice
$$
\alpha = {1\over 3}~.
$$

This consideration strongly suggests the existence of the whole
"brane" $q^+$ action yielding the full Nambu-Goto action \p{NGact}
in the bosonic sector. Then the field redefinition \p{redef}
shows first terms in the bosonic part of the change of variables
from the central basis in the harmonic $(1,0)\; d=6$ superspace,
where the hypermultiplet is described by the superfield
$q^{+a}(Z,u) = q^{ia}(Z)u^+_i$ subjected to the dynamical constraint
\p{basconstr3}, to the analytic basis where the same hypermultiplet
is represented by the manifestly analytic superfield $q^+_a(\zeta)$
possessing highly nonlinear action the first terms of which
are given by \p{corr}. Such an action, at least for the given
particular case, could provide a viable alternative to the
standard GS-type  Lagrangian description
of superbranes \cite{actions}. It should be a natural generalization
of the Goldstone chiral superfields action of ref. \cite{bg1}.
It is worth mentioning that possible existence of such an
off-shell brane action for hypermultiplet was  anticipated
in \cite{hp} based upon the superembedding considerations.

It is still unclear how to find out the analytic basis form
of the nonlinear coset transformations \p{susy1} - \p{ktr}
which should constitute the underlying symmetries of this
hypothetical action. As eq. \p{redef} shows, even the
$R^{1,9}/ R^{1,5}$ translations $q^{ia}(Z) \rightarrow q^{ia}(Z)
+ c^{ia}$ should contain non-linear terms when realized
on $q^{+a}(\zeta)$ (this transformation of $q^{+a}$
starts with the well-known
isometry of the free $q^+$ action \p{qaction} $q^{+a} \rightarrow
q^{+a} + c^{ia}u^+_i$). Good guiding principle in searching for
the full brane $q^+$ action is the preservation of manifest
invariance under both linearly realized mutually commuting
$SU(2)$ groups,
the harmonic one which is realized as the standard automorphism
group of $(1,0)\;d=6$
SUSY and the "Pauli-G\"ursey" one which acts on the index $a$
of $q^{+a}(\zeta)$. It is funny that this last symmetry
playing an important role in the harmonic
superspace approach comes out in the present framework as a part
of the $D=10$ Lorentz group $SO(1,9)$, on equal footing with
the harmonic $SU(2)$ and the $d=6$ Lorentz group $SO(1,5)$.

Another severe restriction is provided by the
dimensionality considerations which require the second
Lagrangian density in \p{corr} and any further corrections to
it to be of dimension $-4$ (in mass units). This implies,
in particular,
that the next, sixth-order term (if it is needed) should
contain the
appropriate number of $x$ or spinor derivatives on $q$'s
(the geometric dimension of $q^{+a}$ is $-1$). Of course,
all such correction terms should have zero harmonic $U(1)$ charge.

Finally, we note that the second term in \p{corr} can be rewritten
as an integral over the analytic superspace with the
Lagrangian density
$$
\sim (q^{+a}\partial^{\al\beta}q^+_a)
(q^{+b}\partial_{\al\beta}q^+_b)~.
$$
This term can be regarded as the appearance of some composite
analytic
vector vielbein $H^{++[\al,\beta]} \sim
q^{+a}\partial^{\al\beta}q^+_a$ in the analytic derivative
${\cal D}^{++}$.
This could be an indication that the full brane $q^+$ action
is representable as a kind of the $q^+$ action in the background
of $(1,0) \; d=6$ supergravity \cite{{sg},{sok6}}, with some
composite superfield vielbeins built out of $q^{+a}$.
Though it is unlikely that the next correction terms would admit
such a simple representation in the analytic superspace.

In principle, the simple action \p{corr} has a chance to be
the sought brane $q^+$ action without any further
correction terms. This potential possibility is related to the fact
that the {\it full} physical bosonic part of \p{corr}
is non-polynomial in $x$-derivatives of $v^{ia}$ and so can
contain the whole Nambu-Goto action (this
on-shell non-polynomiality emerges from solving the auxiliary fields
equations, like, e.g., in the well-known Taub-NUT example \cite{cmp}).
Such an opportunity would be of course very surprizing.

\vspace{0.4cm}
\noindent{\bf 7. Concluding remarks.} One of the most intriguing
and urgent problems for the future study is the
construction of the full 5-brane extension of the free analytic
$q^+$ action \p{qaction}.
Once such an
action is known, one can pose the question as to what
could be brane extensions of
non-trivial $q^+$ actions with hyper-K\"ahler sigma models
in the bosonic sector. The brane version of \p{qaction}
(if existing) should describe, in the bosonic sector, the
$5$-brane with transverse coordinates $q^{ia}(x)$ parametrizing
a flat target manifold $R^4$ (this amounts to the splitting
$R^{1,9} \rightarrow R^{1,5}\otimes R^4$).
Then an analogous extension
of the action of self-interacting $q^+$ is expected to give
the static gauge action of $5$-brane evolving on some
curved $D=10$ manifold $\sim R^{1,5}\otimes H^4$, $H^4$
being a hyper-K\"ahler manifold.

It is interesting to see whether other known $(1,0)\; d=6$
supermultiplets can play a role of Goldstone ones supporting
a partial spontaneous breaking of higher SUSY. Let us examine,
e.g., abelian gauge vector multiplet \cite{hst}.
The fundamental object of $(1,0)\; d=6$ gauge theory is the analytic
harmonic prepotential $V^{++}(\zeta)$ \cite{{gikos},{hsw},{bz}} which
in the WZ gauge collects the components of vector multiplet
in the following suggestive way
\be
V^{++} = \theta^{+\mu}\theta^{+\nu}A_{[\mu \nu]}(x) +
\theta^{+\mu} \theta^{+\nu}\theta^{+\rho}\epsilon_{\mu\nu\rho\lambda}
\psi^{\lambda i}(x)u^-_i + (\theta^+)^4 D^{(ik)}(x)u^-_iu^-_k~.
\label{vect}
\ee
One sees that the physical fermionic field
in this multiplet has the same chirality as $\theta^{\lambda}_i$,
in contrast to the physical fermion in $q^{ia}(x,\theta)$ which
has the
opposite chirality. Thus, if we wish to utilize the $d=6$
vector multiplet as a Goldstone one describing partial breaking
of some higher SUSY, the spontaneously broken and unbroken
spinor generators of the latter should have
the same chirality. In
other words, this multiplet is suitable to represent the partial
supersymmetry breaking $(2,0)\; d=6 \rightarrow (1,0) \; d=6$. In
this notation, the breaking associated with $q^{ia}(x,\theta)$
as the Goldstone multiplet corresponds to the pattern
$(1,1)\; d=6 \rightarrow (1,0) \; d=6$. By analogy with the results
of \cite{bg2}, one can expect that the theory of the vector Goldstone
$d=6$ multiplet is manifestly $(1,0)$ supersymmetric
$d=6$ Born-Infeld theory with hidden nonlinearly realized
$(2,0)$ SUSY. In the brane language, such a theory should
correspond to D5-superbrane. After reduction to $d=4$ the relevant
action should produce $N=2\; d=4$ Born-Infeld action with hidden
$N=4$ SUSY \footnote{An $N=2$ superfield extension of
Born-Infeld action has been recently constructed in \cite{ske}.}.

One more possible candidate for the Goldstone multiplet is the
$d=6$ self-dual tensor multiplet with the following on-shell content
\cite{{bsvp},{sok6}}
\be
\sigma(x)~,\;\;\; B^\alpha_\beta(x)\; (B^\alpha_\beta = 0)~, \;\;\;
\psi_{\alpha i}~.
\label{tens}
\ee
It is capable to support the breakdown of some kind of
$(1,1)\;d=6$ SUSY down to $(1,0)\;d=6$, like the Goldstone hypermultiplet.
It is known to be associated with the PBGS pattern $N=1\;D=7 \;
\rightarrow
\;(1,0)\; d=6$ \cite{{hs},{goteb}}.

Some additional possibilities arise upon the reduction to $d=5$ and
$d=4$.

It is also interesting to study other versions of partial
spontaneous breaking of $N=1\; D=10$ SUSY within this framework.
If we limit our attention
to the $1/2$ breaking, a simple analysis shows that
only one self-consistent option is possible, besides the one
considered
here. It corresponds to breaking $N=1 \;D=10$ SUSY down to
$(8,0)$ (or, equivalently, $(0,8)$) $d=2$ SUSY. From the brane
standpoint, it should provide $(8,0)\;d=2$ worldsheet superfield
PBGS description of the heterotic $N=1\; D=10$ superstring in a flat
background. All other possible $1/2$ SUSY breaking patterns can be
ruled out on the physical grounds: in all of them some
unbroken supersymmetry
generators yield in their anticommutator broken translation
generators, that is in conflict with the interpretation of
these spinor charges as belonging to the vacuum stability subgroup.
It is curious that such simple algebraic PBGS reasonings distinguish
just two self-consistent BPS $N=1\; D=10$ super $p$-branes,
viz., super $5$-brane and superstring.

As one more remark, it is noteworthy that the superfield PBGS approach
can be successfully extended to the most interesting case of
$N=1\; D=11$ (Type IIA $N=2\;D=10$) SUSY. The $(1,0) \; d=6$ Goldstone
superfield framework is well adapted for decription of
the $1/4$ partial
breaking of this SUSY down to $(1,0)\;d=6$ (or $(0,1)\; d=6$) SUSY.
It turns out that in this case the set of fundamental unremovable
Goldstone superfelds is reducible:, besides $q^{ia}(x,\theta)$ it
includes two more superfields. These are a bosonic scalar superfield
$\Phi (x, \theta)$ (it parametrizes
spontaneously broken 11th direction in $R^{1, 10}$) and a fermionic
one $\xi^{a\rho}(x,\theta)$ (it is associated with one of two extra
spontaneously broken $d=6$ supercharges present in $D=11$
SUSY algebra in the $d=6$ notation). The dynamical and irreducibility
constraints on these superfields following from the
$D=11$ SUSY counterparts
of the basic covariant constraint \p{basconstr}, as well as
the brane interpretation of the emerging system will be
presented elsewhere.

Finally, we wish to point out that it is desirable to further clarify
the relationship between the PBGS and superembedding approaches
\footnote{Some aspects of interplay between these two approaches
are touched upon in
one or another way in recent papers \cite{ren} - \cite{abkz}.}.
It seems that the appropriate PBGS description exists for most of
superbranes and it corresponds to choosing the static gauge with respect
to local symmetries in the GS formulation, including $\kappa $-symmetry.
It is a rather difficult task to find such symmetries and
to prove, e.g., invariance of the relevant GS-type actions under them.
At the same time, the PBGS approach deals with a minimal
set of worldvolume superfields accomodating the superbrane
physical degrees of freedom and provides a systematic way
to deduce their transformation laws both under manifest and hidden
symmetries. In a number of cases it also gives precise recipes or,
at least,
hints of how to construct the relevant manifestly worldvolume
supersymmetric off-shell actions. In this sense it should be
regarded as
complementary to the superembedding approach. On the other hand, the
power of the latter consists, in particular, in providing
a possibility to classify all possible physical worldvolume
supermultiplets
related to various superbranes and to learn whether they are on- or
off-shell as a result of imposing some basic constraints on the
relevant
superfields. Actually, all the minimal Goldstone supermultiplets
appearing in the PBGS constructions known so far are in the list of
physical superbrane worldvolume multiplets obtained by
the linearized level analysis of the superembedding equations
in \cite{hs}. In particular, for the $N=1\; D=10$ Type I
super 5-brane
it picks out the hypermultiplet as such a physical multiplet and
predicts it to be on-shell in a precise correspondence with our
PBGS analysis.

\vspace{0.4cm}
\noindent{\bf Acknowledgements.} We thank F. Delduc, R. Kallosh,
S. Ketov, O. Lechtenfeld, A. Pashnev, M. Tonin, M. Vasiliev and,
especially, D. Sorokin for their interest in the work and
illuminating discussions.
E.I. is grateful to Organizers of Conferences in
Dubna and Buckow for giving him a possibility to present
this talk. He also thanks O. Lechtenfeld for the hospitality
at the Institute of Theoretical Physics in Hannover where this
work was finalized. This research was supported in part by the
Fondo Affari Internazionali Convenzione Particellare INFN-Dubna.
The work of E.I. and S.K. was partly
supported by grants RFBR 96-02-17634, RFBR-DFG 96-0200180,
INTAS-93-127ext, INTAS-96-0538 and INTAS-96-0308.

\end{document}